\documentstyle[iopconf1]{article}

\def\be{\begin{equation}}
\def\ee{\end{equation}}
\def\bea{\begin{eqnarray}}
\def\eea{\end{eqnarray}}

\def\ma{m_{\rm a}}
\def\fa{f_{\rm a}}

\def\dalemb#1#2{{\vbox{\hrule height .#2pt
\hbox{\vrule width.#2pt height#1pt \kern#1pt\vrule width.#2pt}
\hrule height.#2pt}}}

\def\tdot{\kern -8.5pt{}^{{}^{\hbox{...}}}}
\def\dotprime{\kern -8.0pt{}^{{}^{\hbox{.}~\prime}}}

\def\lapp{\hbox{$ {     \lower.40ex\hbox{$<$}
                   \atop \raise.20ex\hbox{$\sim$}
                   }     $}  }
\def\gapp{\hbox{$ {     \lower.40ex\hbox{$>$}
                   \atop \raise.20ex\hbox{$\sim$}
                   }     $}  }

\def\marbul{\strut\vadjust{\kern-2pt$\bullet$}}

\mathchardef\less="321C

\def\smallskip{\vskip 10pt}
\def\microeV{\mu\hbox{\rm eV}}

\begin{document}
\title{Axion string cosmology and its controversies}

\author{R.A. Battye and E.P.S. Shellard~\dag}

\affil{\dag\ Department of Applied Mathematics and Theoretical
Physics, University of Cambridge, \ Silver Street, Cambridge CB3 9EW,
U.K.}

\beginabstract

Understanding axion cosmology has important experimental consequences
since it constrains the range of allowed values for the axion mass.
In the standard thermal scenario, which assumes Peccei-Quinn symmetry
restoration after inflation, an axion string network forms at the
phase transition $T\sim \fa$ and then radiatively decays into a
cosmological background of axions. Under standard assumptions for the
evolution of this string network and the radiation from it, axions
must have a mass $\ma \sim 100\,\mu$eV with specified large
uncertainties. We discuss critically the various suggestions in the
literature that the axion mass might be lighter. 

\endabstract

\section{Introduction}

The axion has remained a popular dark matter candidate because of its
enduring motivation as an elegant solution to the strong CP-problem~\cite{PecQui77}.  Despite early hopes of discovery, 
it turned out that in order to be consistent with accelerator
searches and astrophysics, the axion must be nearly `invisible' and extremely 
light. Its couplings and mass are inversely proportional to the (large)
Peccei-Quinn (PQ) scale $\fa$ as in
\be
\ma = 6.2 \times 10^{-6}\hbox{eV} \left ( 10^{12}\hbox{GeV}\over \fa\right)\,.
\label{axionmass}
\ee
Accelerator constraints have been largely superseded by those from 
astrophysics; because the axion is so weakly coupled, volume effects can 
compete with other surface and convective stellar energy loss mechanisms.
The strongest astrophysical constraints on 
the axion mass derive from studies of supernova 1987a and conservative
estimates yield
$\ma ~\lapp~ 10\,\hbox{meV}$~\cite{Raf97}.  The present programme of 
large-scale axion search experiments~\cite{Hagetal98} are
sensitive to a  mass range $m_a \sim 1\hbox{--}10\,\microeV$, which has been 
chosen for a variety of historical and technological reasons. Here, we
discuss the motivation from cosmology for a larger mass axion
($\ma\sim 100\mu{\rm eV}$) if it
forms a substantial part of the dark matter of the universe. We will
also comment on suggestions in the literature that question this point
of view.

\section{Standard axion cosmology}

\noindent The cosmology of the axion is determined by the two 
energy scales $\fa$ and $\Lambda_{\rm QCD}$.  The first important event
is the PQ phase transition which is broken at a high temperature
$T\sim \fa \gapp 10^9$GeV.  This creates the axion,
at this stage an effectively massless pseudo-Goldstone boson, 
as well as a network of axion 
strings~\cite{VilEve82} 
which decays gradually into a background cosmic axions~\cite{Dav86}.
(Note that one can engineer models in which an 
inflationary epoch interferes with the effects of 
the PQ phase transition, this possibility is discussed in
ref.~\cite{SheBat98}).
At a much lower temperature $T\sim \Lambda_{\rm QCD}$
after axion and string formation, instanton effects `switch on', the axions
acquire a small mass, domain walls form~\cite{Sik82} between the 
strings~\cite{VilEve82} 
and the complex
hybrid network annihilates into axions in about one Hubble time~\cite{She86}.  

There are three possible mechanisms by which axions are produced in the
`standard thermal scenario', which assumes PQ symmetry restoration
after any inflationary epoch:  (i) thermal production, (ii) axion string
radiation and (iii) hybrid defect annihilation when $T=\Lambda_{QCD}$.  
Axions consistent with the astrophysical bounds must decouple from 
thermal equilibrium very early;
their subsequent 
history and number density is analogous to the decoupled neutrino,
except that unlike a 100eV massive neutrino, thermal axions cannot hope 
to dominate the universe with $\ma\lapp 10$meV.  
We now turn to the two dominant axion production mechanisms, but first we 
address an important historical digression.

\subsection{Misalignment misconceptions} 
The original papers on axions suggested 
that axion production
primarily occurred, not through the above mechanisms, but 
instead by `misalignment' effects at the QCD phase transition
~\cite{PAD83}.  Before the
axion mass `switches on', the axion field $\theta$ takes random 
values throughout
space in the range 0 to $2\pi$; it is the phase of the PQ-field lying at the 
bottom of a $U(1)$ `Mexican hat' 
potential.  However, afterwards the potential becomes tilted and 
the true minimum 
becomes $\theta=0$, so the field in the `misalignment' picture begins to 
coherently oscillate about this minimum; this homogeneous mode corresponds
to the `creation' of zero momentum axions. Given an initial
rms value $\theta_{\rm i}$ for these oscillations, it is 
relatively straightforward to 
estimate the total energy density in zero momentum axions and compare 
these to the present mass density of the universe (assuming
a flat $\Omega=1$ FRW cosmology)~\cite{PAD83,Tur86}:
\be
\Omega_{{\rm a},{\rm hom}}  ~\approx ~
 2 \,\Delta h^{-2}\;\theta_{\rm i}^2 
{\rm f}(\theta_{\rm i})\,
\bigg{(}{10^{-6}{\rm eV}\over \ma}\bigg{)}^{1.18}  
\label{homcount}
\ee
where $\Delta\approx 3^{\pm 1}$ accounts for both model-dependent axion 
uncertainties and those due to the nature of the QCD phase 
transition, and $h$ is the rescaled Hubble parameter. 
The function ${\rm f} (\theta)$ is an anharmonic correction for 
fields near the top of the potential close to  unstable 
equilibrium $\theta\approx\pi$, that is, 
with ${\rm f}(0)=1$ at the base $\theta\approx 0$ and diverging 
logarithmically 
for $\theta\rightarrow\pi$~\cite{SheBat98}.
If valid, the estimate (\ref{homcount}) would imply a 
constraint $\ma\gapp 5\mu {\rm eV}$ for the anticipated
thermal initial conditions with $\theta_{\rm i} = 
{\cal O}(1)$~\cite{PAD83,Tur86}. 

As applied to the thermal scenario, the expression (\ref{homcount}) 
is actually a very considerable
underestimate for at least two reasons:  First, the axions are not `created' by 
the mass `switch on' at $t=t_{QCD}$, they are already there 
with a specific
momentum spectrum $g(k)$ determined by dynamical mechanisms prior to this time.
The actual axion number obtained from $g( k)$ is much 
larger than the rms average assumed in (\ref{homcount}) 
which ignores the true particle content.  Secondly,
this estimate was derived before much stronger topological 
effects were realized, notably the presence of 
axion strings and domain walls. In any case, these
nonlinear effects complicate the oscillatory behaviour considerably,
implying that the homogeneous estimate (\ref{homcount}) is poorly motivated.

\subsection{Axion string network decay} 

Axions and axion strings are inextricably intertwined.  Like ordinary
superconductors or superfluid $^4$He, axion models contain  a broken
$U(1)$-symmetry and so there exist vortex-line solutions.  Combine
this fact with the PQ phase transition, which means the field is
uncorrelated beyond the horizon, and a random network of axion strings
must inevitably form.  An axion string corresponds to a non-trivial
winding from $0$ to $2\pi$ of the axion field $\theta$ around the
bottom of its  `Mexican hat' potential.  It is a global string with
long-range fields, so its energy per unit length $\mu$ has a
logarithmic  divergence which is cut-off by the string curvature
radius $R\lapp t$, that is, $\mu \approx 2\pi \fa^2 \ln (t/\delta)\,,$
where the string core width is $\delta \approx \fa^{-1}$.  The axion
string, despite this logarithmic divergence,  is a strongly localized
object and is likely to behave as such; if we have a string stretching
across the horizon at the QCD  temperature, then $\ln(t/\delta)\sim
65$ and over 95\% of its  energy lies within a tight cylinder
enclosing only 0.1\% of  the horizon volume.  To first order, then,
the string behaves like  a local cosmic string, a fact that can be
established by a precise analytic derivation and careful comparison
with numerical  simulations~\cite{BatShe94a}.  

After formation and a short period of damped evolution, the axion string
network will evolve towards a scale-invariant regime with a fixed number of
strings crossing each horizon volume (for a cosmic string 
review see ref.~\cite{VilShe94}). 
This gradual demise of the network
is achieved by the production of small loops which oscillate relativistically 
and radiate primarily into axions.
The overall density of strings splits neatly into
two distinct parts, long strings with length $\ell > t$ and a population of 
small loops $\ell < t$, that is, $\rho=\rho_{\infty}+\rho_{L}$.
High resolution numerical simulations confirm 
this picture of string 
evolution and suggest that the long 
string density during the radiation era is $\rho_\infty \approx 
13\mu/t^2$~\cite{BBAS90}. 
To date, analytic descriptions of the loop distribution have used the 
well-known string `one scale' model, which predicts a
number density of loops defined as $\mu\ell\,n(\ell,t)\,d\ell=\rho_{L}
(\ell,t)d\ell$ in the interval $\ell$ 
to $\ell+d\ell$ to be given by 
\be
n(\ell,t)={4\alpha^{1/2}(1+\kappa/\alpha)^{3/2}
\over (\ell+\kappa t)^{5/2} t^{3/2}}
\,,\label{numdenloop}
\ee 
where $\alpha$ is the typical loop creation size relative to the horizon 
and $\kappa \approx 65/[2\pi \ln (t/\delta)]$ is the loop radiation 
rate~\cite{BatShe94b}. Once formed at $t=t_0$ with length $\ell_0$, a typical loop 
shrinks linearly as it decays into axions $\ell = \ell_0 - \kappa(t -t_0)$.
The key uncertainty in this treatment is the loop creation size $\alpha$, but 
compelling heuristic arguments place it near the radiative backreaction scale, 
$\alpha \sim \kappa$.  (If this is the case, we note that the loop contribution 
is over an order of magnitude larger than direct axion radiation from long
strings.)

String loops oscillate with a period $T=\ell/2$ and radiate
into harmonics of this frequency (labelled by $n$), just like other 
classical sources.  Unless a loop has a particularly 
degenerate trajectory, it will have a radiation spectrum $P_n \propto n^{-q}$
with a spectral index $q>4/3$, that is, the spectrum is dominated by 
the lowest available modes. Given the loop density (\ref{numdenloop}), we can 
then calculate the spectral number density 
of axions $dn_{\rm a}/d\omega$,
which turns out to be 
essentially independent of the exact loop radiation spectrum 
for $q>4/3$.   From this expression we can integrate over $\omega$
to find the total axion number at the time $t_{\rm QCD}$, that is, when the axion 
mass `switches on' and the string network annihilates. Subsequently,
the axion number will be conserved, so we can find the number-to-entropy ratio
and project forward to the present day.  Multiplying the present number
density by the axion 
mass $\ma$ yields the overall axion string contribution to the density of 
the universe~\cite{BatShe94b}:
\be
\Omega_{\rm a,string}\approx 110\Delta h^{-2}
\bigg{(}{10^{-6}\hbox{eV}\over \ma}\bigg{)}^{1.18}f(\alpha/\kappa)
\,,\label{stringbound}
\ee
where
\be
f(\alpha/\kappa)=\bigg{[}\bigg{(}1+{\alpha\over\kappa}\bigg{)}^{3/2}-1\bigg{]}\,.
\ee
The key additional 
uncertainty from the string model is the ratio $\alpha / \kappa\sim {\cal O}(1)$,
which  
should be clearly distinguished from 
particle physics and cosmological uncertainties inherent in $\Delta$ and $h$
(which appear in all estimates of $\Omega_{\rm a}$).  With a Hubble parameter 
near $h=0.5$,
the string estimate (\ref{stringbound}) tends to favour a dark matter axion with a mass 
$\ma \sim 100\mu$eV, moreover
a comparison with (\ref{homcount}) confirms that
$\Omega_{\rm a,string}$ is well over an order of magnitude 
larger than the `misalignment' contribution.

\subsection{Hybrid defect annihilation}
Near the QCD phase transition the axion acquires a mass and network
evolution alters dramatically because domain walls form. 
Large field variations around the strings collapse into 
these domain walls, which subsequently begin to dominate over 
the string dynamics.  This occurs when the wall 
surface tension $\sigma$ becomes comparable to the 
string tension due to the typical curvature $\sigma\sim\mu/t$. 
The demise of the hybrid string--wall network proceeds rapidly, as
demonstrated numerically~\cite{She86}.  The strings frequently intersect and
intercommute with the walls, effectively `slicing up' the network into small
oscillating walls bounded by string loops.  Multiple self-intersections will reduce
these pieces in size until the strings dominate the dynamics again and decay
continues through axion emission.

An order-of-magnitude estimate of the demise of the string--domain wall
network indicates that there is an additional contribution~\cite{Lyt92} 
\be
\Omega_{\rm a,dw}\sim{\cal O}(10)\Delta h^{-2}
\bigg{(}{10^{-6}\hbox{eV}\over \ma}\bigg{)}^{1.18}\,.
\ee
This `domain
wall' contribution is ultimately due to loops which are created at 
the time $\sim
t_{\rm QCD}$. Although the resulting loop density will be similar 
to (\ref{numdenloop}),
there is not the same accumulation from early times, so it is likely to be
subdominant~\cite{BatShe94b} relative to (\ref{stringbound}). 
More recent work,~\cite{Nag97}
questions this picture by suggesting that the walls stretching between 
long strings dominate and will produce a contribution anywhere in 
 the wide range $\Omega_{\rm a,dw} 
\sim (1\hbox{--}44) \Omega_{\rm a,string}$; however, this assertion 
requires stronger quantitative support.
Overall, like most effects,\footnote{We note briefly that it is also 
possible to weaken any axion mass bound through 
catastrophic entropy production between the QCD-scale and nucleosynthesis,
that is, in the timescale range $10^{-4}s \lapp t_{\rm ent} \lapp 10^{-2}s$.  
Usually this involves the energy density of 
the universe becoming temporarily dominated by an exotic massive particle 
with a tuned decay timescale.} the domain wall
contribution will serve to further strengthen the string  
bound (\ref{stringbound}) on the axion.

Up to this point we have only considered the simplest axion models
with a unique vacuum $N=1$, so what happens when $N>1$?  In this case,
any strings  present become attached to $N$ domain walls at the
QCD-scale.  Such a  network `scales' rather than annihilates, and so
it is cosmologically disastrous being incompatible (at the very least)
with CMB  isotropy.

\section{Theoretical challenges to $\ma\sim 100\mu{\em eV}$}

\noindent The conclusion that that $\ma$ is most likely around
$100\mu{\rm eV}$ is an important one since the only experimental
search for axions with a realistic level of sensitivity can only
detect axions if  $\ma\sim 1-10\mu{\rm eV}$. The possibility of
detection by the current generation of experiments might still be
possible, either by pushing the theoretical and cosmological
uncertainties of the  calculation presented in the previous section to
the limit, allowing for entropy production by a late decaying massive
particle,  or by allowing an inflationary epoch which prevents PQ
symmetry restoration\footnote{We should note that as discussed in
ref.~\cite{SheBat98}, even if the inflationary reheat temperature is
below $\fa$, it is still possible for the strings to be formed and for
them to be the dominant source of axions.}. However, the theoretical
framework which we have presented in the previous section has been
also challenged conceptually and quantitatively in the literature and
here we shall comment critically on this work.

The basis of our understanding global string evolution and radiation
presented in the previous section is based on the assumption that
global strings behave in an essentially similar way to local strings
described by the an extension of the Nambu action, known as the
Kalb-Ramond action. This kind of action is used frequently in
superstring theory to describe the dynamics of strings coupled to a
massless scalar field, and its efficacy for describing global string
dynamics was established  quantitatively  in
ref.~\cite{BatShe94a}. Since the global strings have long range forces
and a logarithmically divergent energy density, one might wonder how
this is possible. But as we have already pointed out over 95\% of the
energy lies within 0.1\% of the volume for cosmological scales. For
the point of view of this discussion we will separate the challenges
in the literature into the conceptual~\cite{Sik} and the
quantitative~\cite{Yama}. Both are just supported by numerical simulations, rather than the analytic arguments and numerical
calculations used in our description. The main point being that
use of numerical simulations of limited physical size and dynamic
range can be hazardous when making extrapolations to cosmological
scales where the strength of the relevant effects is very much diminished.

\subsection{Conceptual issues : the spectrum of radiation from
strings}

\noindent While the issue of the spectrum of radiation from global
strings is a rich and interesting theoretical subject, its effects on
the cosmological axion density are much simpler than often
realised. The main issue, as elucidated in ref.~\cite{BatShe98},
relates to the power in high frequency modes. The frequency of the
$n$th radiation mode for a loop is $\omega_n=4\pi n/l$, where $l$ is
its length. If the fall-off of the power spectrum of radiation is a
power law $(P_n\propto n^{-q}$ for large $n$, as suggested by
analytic calculations using the Kalb-Ramond action~\cite{VilVach}, then so long
as $q>1$ the calculation we presented in the previous section is
still valid. This is because only in the divergent case $q=1$
does the calculation of the axion density depend sensitively on the
high frequency cut-off~\cite{Sik}. A wide range of analytic and numerical
calculations support this issue, but since this issue has been
discussed extensively in a large number of publications, we see no
reason to discuss it further here.

\subsection{Quantitative issues : the scaling density of strings}

\noindent Recent simulations of a network of global strings on a
`large' ($256^3$) grid has suggested that the scaling density of
strings $\xi=\rho_\infty t^2\mu\approx 1$, rather than $\xi\approx 13$
as used in our calculation. This clearly has a significant effect on
the axion mass prediction, reducing it to around $10\mu{\em eV}$,
since the axion density depends linearly on $\xi$. The argument that
$\xi\approx 13$ is based on the premise that the cosmological scale
global strings behave in an essentially similar way to the Nambu
strings for the reasons already discussed, and this may have to be
modified  to account for any alight deviations from this.  However, in
the simulations presented in ref.~\cite{Yama} there is no reason to
believe that the strings will behave anything like Nambu strings
because $\log(t/\delta)$ is at most 4, that is, in the same 0.1\% of
the volume of space, there is less than 50\% of the energy. The
long range forces between strings  will be more than 10 times stronger
than in the cosmological context and will substantially effect the
dynamics and hence the scaling density.

Furthermore, the dynamic range of the simulations is much smaller than
it appears. The time of formation of the string network is
$t_{\rm f}=20t_{\rm i}$, where $t_{\rm i}$ is the start of the
simulation, and the time of the end of the simulation is $t_{\rm
e}=80t_{\rm i}$. Hence, in real time $t$ the dynamic range is $t_{\rm
e}/t_{\rm f}=4$, not 80 as one might think, and the true dynamic range
in conformal time is only $\tau_{\rm e}/\tau_{\rm f}=(t_{\rm e}/t_{\rm
f})^{1/2}=2$. Therefore, the simulations are too small to simulate
realistic cosmological behaviour and have very limited dynamic range.


\def\jnl#1#2#3#4#5#6{\hang{#1, {\it #4\/} {\bf #5}, #6 (#2).}}


\def\jnlerr#1#2#3#4#5#6#7#8{\hang{#1, {\it #4\/} {\bf #5}, #6 (#2).
{Erratum:} {\it #4\/} {\bf #7}, #8.}}


\def\jnltwo#1#2#3#4#5#6#7#8#9{\hang{#1, {\it #4\/} {\bf #5}, #6 (#2);
{\it #7\/} {\bf #8}, #9.}}

\def\prep#1#2#3#4{\hang{#1 (#2),  #4.}}

\def\myprep#1#2#3#4{\hang{#1 (#2), '#3', #4.}}

\def\proc#1#2#3#4#5#6{\hang{#1 (#2), `#3', in {\it #4\/}, #5, eds.\ (#6).}
}
\def\procu#1#2#3#4#5#6{\hang{#1 (#2), in {\it #4\/}, #5, ed.\ (#6).}
}

\def\book#1#2#3#4{\hang{#1 (#2), {\it #3\/} (#4).}
									}

\def\genref#1#2#3{\hang{#1 (#2), #3}
									}


\def\prl{Phys.\ Rev.\ Lett.}
\def\pr{Phys.\ Rev.}
\def\pl{Phys.\ Lett.}
\def\np{Nucl.\ Phys.}
\def\prp{Phys.\ Rep.}
\def\rmp{Rev.\ Mod.\ Phys.}
\def\cmp{Comm.\ Math.\ Phys.}
\def\mpl{Mod.\ Phys.\ Lett.}
\def\apj{Ap.\ J.}
\def\apjl{Ap.\ J.\ Lett.}
\def\aap{Astron.\ Ap.}
\def\cqg{Class.\ Quant.\ Grav.} 
\def\grg{Gen.\ Rel.\ Grav.}
\def\mn{M.$\,$N.$\,$R.$\,$A.$\,$S.}
\def\ptp{Prog.\ Theor.\ Phys.}
\def\jetp{Sov.\ Phys.\ JETP}
\def\jetpl{JETP Lett.}
\def\jmp{J.\ Math.\ Phys.}
\def\cupress{Cambridge University Press}
\def\pup{Princeton University Press}
\def\wss{World Scientific, Singapore}

\end{document}